\documentclass[dvips]{article}
\usepackage{graphics}

\include{macros}

\begin{document}

\title{Two-band second moment model for transition metals and alloys}

\author{ Graeme J.~Ackland}

\date{}

\maketitle
\vspace*{-1cm}
\begin{center}
{\small
Centre for Science at Extreme Conditions,
School of Physics and Astronomy, The University of Edinburgh, \\
James Clerk Maxwell Building, The King's Buildings,\\
Mayfield Road, Edinburgh EH9 3JZ, UK

}\end{center}

\bibliographystyle{unsrt}

\section{Abstract}

A semi-empirical formalism based on the second moment tight binding
approach, considering two bands
  is presented for deriving interatomic potentials for
magnetic d-band materials.  It incorporates an empirical local
exchange interaction, which accounts for magnetic effects without
increasing the computing time required for force evaluation.  
The consequences of applying a two-band picture to transition metal alloys and transition metal impurities is examined, which combined with evidence from {\it ab initio calculations} leads to some surprisingly simplifying conclusions.

{\it PACS  61.80.Az 71.20.Be 61.82.Bg 71.15.-m}

\section{Introduction}

Semiempirical models for metallic binding have had a long and
successful history in computer modelling.  The distinguishing features
of models for metallic bonding are that they are short-ranged and
non-pairwise.  The first of these features arises from the strong
screening of the nuclear charge by the mobile electrons, the second
arises from the delocalisation of those electrons.

The most significant development in accounting for many body effect came
in the mid eighties with the implementation of `embedded atom'
potentials (EAM)\cite{db} (based loosely on density functional
theory\cite{HKS}) and `Finnis-Sinclair' potentials (FS)\cite{FS} (based on
the tight binding second moment approximation\cite{duc}).  The two
models have very similar computational requirements, and the names are
often used interchangably, however there are some distinctions which
come to the fore when considering multicomponent alloys.

To highlight the differences, the energy according to the EAM is written:

\begin{equation} 
U_{EAM} = \sum_{ij} V_{IJ}(r_{ij}) + \sum_i F_I[\sum_j \phi_J(R_{ij}) ]
\end{equation}

Where $i$ and $j$ label atoms of element $I$ and $J$ respectively, 
$V$ is a pairwise potential which depends on both species, $F_I$ and $\phi_J$ 
are the embedding function and charge density which depend on one species only.

In slight contrast, the FS approach implies:
\begin{equation} U_{FS} = \sum_{ij} V_{IJ}(r_{ij}) + \sum_i F[\sum_j \phi_{IJ}(R_{ij})] \end{equation}

where $\phi_{IJ}$ is the squared hopping integral between the atoms on site 
$i$ and site $j$ and F is independent of the atomic species 
($F(x)=\sqrt{x}$ in the second moment tight binding approximation)
In relating theses potential to tight-binding, there is a subtle issue 
of interpretation whether the many body part is a {\it bond} or {\it band} 
energy\cite{PettiforBook} and extensions beyond second moment work 
explicitly with bonds\cite{bop}.  While mindful of the distinction, 
for convenience the discussion here is presented in terms of bands projected onto atoms.

This subtle distinction becomes relevant for alloys: under EAM one
requires no further refitting of the many body function for each
species, while the FS implies that a new $\phi_{IJ}$ 
function should be fitted for each pair of elements.

The success of {\it ab initio} methods in describing materials
behaviour, and the possibility of using it directly to parameterise
kinetic Monte Carlo simulations, may appear to negate the need for
empirical potentials.  However the lengthscales of correlated events
involved in certain types of calculation - such as radiation damage cascades from
fusion neutrons - are so large that empirical potentials still have a 
role to play.

{\it ab initio} calculation has revealed a number of shortcomings in
previous parameterisations of interatomic potentials for radiation
damage application.  In particular, when the original EAM
parameterisation were done in the 1980s, the only information which
affected the fit of the potential at short range came from isotropic
compression: extrapolation of fits to empirical equations of state
such as Murnaghan or Rose\cite{rose}, or simple quantum electron gas
calculations\cite{thet}.  Recent total energy calculations reveal a
picture in which isotropic compression of metals is resisted by
many-body effects (such as electron kinetic energy), and that ions can
approach one another rather more closely than had been predicted.  A
consequence of this is typically that interstitial formation energies
are lower than in previous parameterisations had predicted, and the 
associated strain fileds are smaller\cite{SuengWu}.

A philosophical issue arises at this stage.  It is clear that such
many body potentials cannot contain all the correct physics of the
systems they are describing.  What then should they aim to achieve?

In early work, it was attractive to show how the physical anomalies of
pair potentials (zero Cauchy pressure, equivalence of vacancy
formation and cohesive energy) can be eliminated with relatively
little computational cost.  In nuclear materials simulation however,
the emphasis is less on physical elegance and more on practical
application.  In the multiscale modelling
framework molecular dynamics modelling is simply an interpolation tool
between exact {\it ab initio} and experimental results and the
non-equilibrium behaviour of many-particle systems. Thus there is
little justification for constraining the ability to precisely
reproduce defect properties on the grounds of physical elegance.  Consequently,
recent parameterisations of ``many-body'' (MB) models write the energy in
the most general form:

\begin{equation}U_{MB} = \sum_{ij} V_{IJ}(r_{ij}) + \sum_i F_I[\sum_j \phi_{IJ}(R_{ij})] \label{eq:MB} \end{equation}

which for an N-component systems gives $N(N+1)$ functions available for 
fitting.

Most of the physical discussion motivating the potential centres on
the form of the many body term, the pairwise part representing
``everything else''.  However, for high energy collisions and
interstitials it is the pairwise part which tends to control behaviour.

It is possible to generate still more complex potentials with a
similar computational cost to EAM.  Examples include the MEAM\cite{Baskes}, which
incorporates angular effects in a way which uses only quantities
already calculated in central-force molecular dynamics models, and the
two-band model which describes the electronic state of each atom in an
analytically solvable form\cite{skr}.

One important system which is not easily described by these methods is
iron.  Although there have been a number of parameterizations for iron
in a single phase, the essential physics leading to the stabilisation of
the bcc phase is missing.  Ab initio calculations show that without
ferromagnetism, the fcc structure is favoured, and that the transition
to the high pressure (hcp) or high temperature (fcc)  phases is
accompanied by a loss of ferromagnetism. Similarly, the atoms close to 
interstitial defects have a reduced magnetisation\cite{domain1}, which 
has the effect of making them more compressible.

In a recent paper\cite{skr}, a simple two-band picture was derived for
Cs to capture s-d electron transfer.  This described the second-moment
tight binding energy in terms of a single parameter (the transfer of
electrons between bands) at each site. Surprisingly, the parameterization proved transferrable across the transition metal series.  Crucially for use in molecular
dynamics, the minimum-energy value of this parameter can be evaluated
at each atom independently and analytically, so the potential has the
same computer requirement as EAM.  Moreover, the energy is variational
in this parameter, meaning that forces take a very simple form.

Here it is shown how a similar
two-band picture within the second-moment approximation can be
developed to incorporate band ferromagnetism.  In the simplest
approximation, it give the surprising result that two-band effect can
be captured in a standard many-body parameterisation, only the interpretation 
of the embedding function being changed.

\section{Two d-band model - magnetism}

\subsection{Modified Many-body term}
In this section we show that incorporating band magnetism in the FS
picture while ignoring the repulsion effect of enhanced Pauli exclusion gives
rise to an embedded atom-type formalism with a non-monotonic embedding
function.

In the second moment approximation to tight binding, the cohesive energy
is proportional to the square root of the bandwidth, which can 
be approximated as a sum of pairwise potentials representing squared 
hopping integrals\cite{FS}.  Assuming atomic charge neutrality, this argument 
can be extended to all band occupancies and shapes\cite{avf}.   For simplicity,
consider a rectangular d-band of full width W centred on $E_0$. 
The bond energy  for a single spin band relative to the free atom
is given by:

\begin{equation}  U^\uparrow  = \int_{-W/2}^{E_f=( Z/N-\frac{1}{2})W }  E/W dE = 
 \frac{Z^\uparrow}{N} (\frac{Z^\uparrow}{N}-1)W/2 \end{equation}

where Z is the occupation of the band and uparrows denote 
``spin up''.

To describe the ferromagnetic case, it is assumed that there are two
independent $d$-bands corresponding to opposite spins, and that these can
be projected onto an atom.  In the atomic case Hund's rules lead to a
spin of S=2 for iron, and there is an energy $U^x$ associated with
transferring an electron to a lower spin state.  In the solid, the simplest 
method is to set $U^x$
to be proportional to the spin with the coefficient of
proportionality being an adjustable parameter, $E_0$.

 \begin{equation}U^{X} = -E_0|Z^\uparrow - Z^\downarrow| 
 \label{hund}   \end{equation}

\begin{figure}[htb]
\centerline{\scalebox{0.6}{
\includegraphics{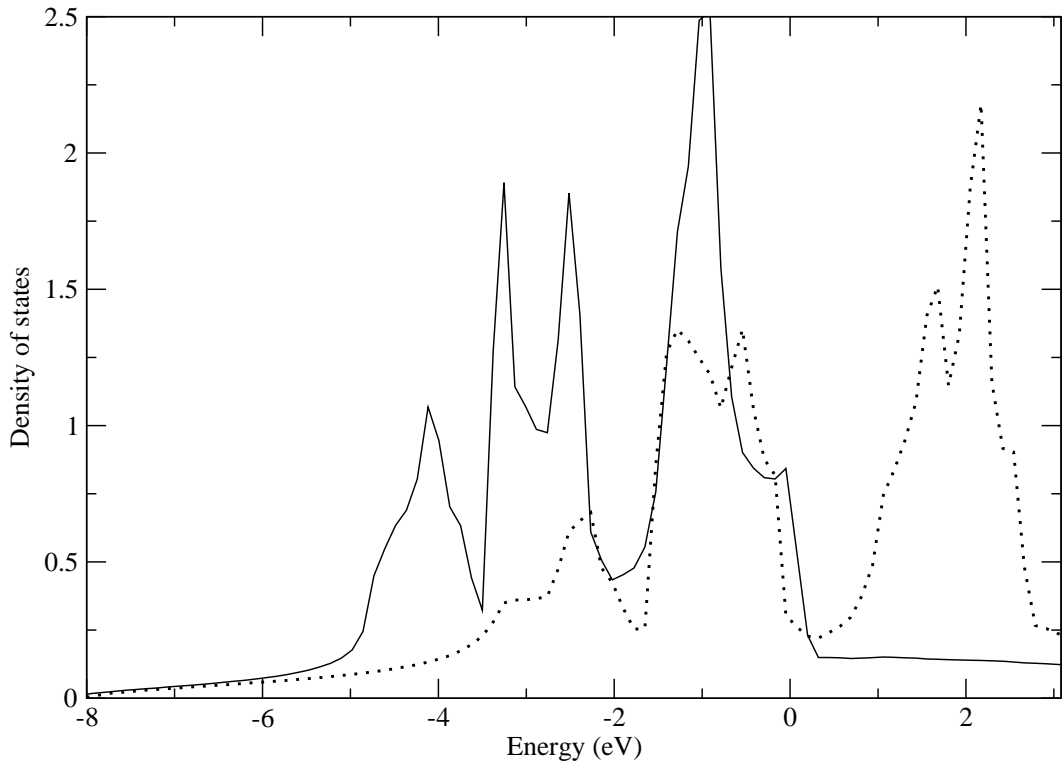}}}
\caption{\label{fig:FeDOS} BCC iron density of states for majority and minority
spin bands from ab initio spin-dependent GGA pseudopotential
calculations with 4913 k=points, adjusted so the ths Fermi energy lies
at the zero of energy. The two band model assumes that the bands have
the same shape and width, but are displaced in energy relative to one
another.  Here it can be seen that this is the case.}
\end{figure}

Defining the spin, $S = Z^\uparrow - Z^\downarrow$ and noticing that
with charge neutrality the total number of electrons at a site is
conserved $T = (Z^\uparrow + Z^\downarrow)$, we make the assumption that the
bandwidth W is the same for each band (see fig.\ref{fig:FeDOS}). The two-band
binding energy\cite{skr} for a d-band with capacity N=5 on a site $i$ is then:

\begin{equation} U_i = U_i^\uparrow + U_i^\downarrow + U_i^{X} = +
\frac{W_i}{4N}(T^2+S_i^2) - TW_i/2 - E_0S_i  \label{energy1}   \end{equation}

This equation gives us the optimal value for the magnetisation of a
given atom of $S_i = 2NE_0/W_i$, and the many-body energy of an atom 
with T=6, N=5 (suppressing the $i$ label) as:

\begin{eqnarray}
U = &  -6W/5 - 5E_0^2/W &   E_0/W_i \le 0.4   \nonumber \\     
  = &   -2W/5 - 4E_0 &  E_0/W_i \ge 0.4       \end{eqnarray}

Which introduces the constraint that there is a maximum value for $S$. 
For a material with T d-electrons (where $T>5$) .
transfer of electrons between the spin bands becomes
advantageous for $W > 10E_0/(10-T)$.  For smaller $W$ the 
spin $\uparrow$ band is full  and the energy is simply proportional to 
the bandwidth of the $\downarrow$ band as in the Finnis-Sinclair model.

Within the second moment model, the bandwidth W is given by the 
square root of the sum of the squares of the hopping integrals\cite{duc}. 
This sum can be represented by a pairwise potential and is the same for each 
band.  

\begin{equation}
W_i = \sqrt{ \sum_j \phi(r_{ij}) }
\end{equation}

Interatomic potentials also
include the effect of the ions and non-valence electrons: this is
usually represented by a simple pairwise potential.   This gives us the
final functional form for incorporating magnetism in the ferromagnetic regime 
within the Finnis-Sinclair formalism:

\begin{equation} E = \sum_j \left[ \sum_i V(r_{ij})  - \sqrt{\rho_j} - B/\sqrt{\rho_j}H(2W-5E_0) - 4E_0 H(5E_0-2W)
\right ] \end{equation}

where H is the Heaviside step function, B is a constant, 
$\rho_j = \sum_i \phi(r_{ij})$, $V$ and $\phi$ are empirically 
fitted pairwise potentials, and the zero of energy is shifted to 
correspond to the {\it non-magnetic} atom.

Note that this form does not explicitly include $S$, and that it has
the form of the embedded atom model with an embedding function $
F_I(x) = \sqrt{x}(1 - B/x) $.  In particular, the local magnetisation
means that the embedding function now depends on the type of atom
through the amount of the particular band projected onto it.

Although this model incorporates magnetism, and provides a way to
calculate the magnetic moment at each site, it is possible to use it without
actually calculating $S$ - the implementation with an
additional many-body repulsive term is similar to the many-body
potential method of Mendelev {\it et al}\cite{mendel}. In terms of forces and energetics, it will
perform in a similar way.  In particular, there is no first order 
magnetic transition.

The anomaly with this model is that the linear expression for $U^{X}$
leads to a non-zero magnetisation at all densities (although this
becomes vanishingly small at high densities).  If instead of
equation.\ref{hund} we use a Stoner-type quadratic term $U^{X} = I_0S^2$
 this changes the eqn.\ref{energy1} such that the
spin state flips from fully magnetised to fully demagnetised at
$W_i=4NI_0$.
 
Neither of these behaviours is quite correct, and a better
treatment involves making the pairwise term spin-dependent, which is
the subject of the following section.

\subsection{Incorporating Pauli Repulsion}

Pauli repulsion arises from electron eigenstates being orthogonal.
 While its nature on a single atom is complex, its
interatomic effects can be modelled as a pairwise effect of repulsion
between electrons of similar spins.  The secondary effect of
magnetisation is that there are more electrons in one band than
another, more same-spin electron pairs to repel one another, 
and so the repulsion between those bands is enhanced. Using the 
Stonor-type $U^{X} = I_0S^2$,
the many body energy on an atom becomes

\begin{equation} U = U^\uparrow + U^\downarrow + U^{X} = +
\frac{W}{4N}(T^2+S^2) - TW/2 + I_0S^2   \end{equation}

This shows that a high magnetisation changes the  energy in two ways, 
firstly by effectively widening the bands and secondly via $I_0$.

We now consider the pairwise repulsive part of the potential.
Conceptually, this contains two effects, the standard paramagnetic
repulsion which represents the screened Coulomb repulsion of the ions
and the core-core repulsion, and an additional $S_i$ dependent term
arising from Pauli repulsion between like-spin electrons\footnote{ Note there are issues about antiferromagnetism here.  
An AFM state
with $S_i+S_j=0$ would have a lower repulsive energy.  In the
tight-binding picture this would be compensated by a much reduced
hopping integral and hence lower $W$.  If we insist on $S_i>0$ 
then we suppress these solutions and can model ferromagnetic or 
diamagnetic iron.  Also, as with DFT-GGS/LDA the spin is Ising-like.}
.  Writing
this in a separable form:

\begin{equation}V(r_{ij}) = V_0(r_{ij}) + (S_i + S_j) 
V_m(r_{ij})\label{V_sep}\end{equation}

The total energy on the $i^{th}$ atom is then

\begin{equation}
U_{tot} = \sum_i \left[ -TW_i/2 + (S_i^2+T^2)W_i/4N + S_iI 
+ \frac{1}{2}\sum_jV_0(r_{ij})+S_i\sum_j V_m(r_{ij}) \right ]
\end{equation}

The key to this step how we
``assign'' the Pauli energy to each atom.  If we choose to do it such
that the energy assigned to each atom $i$ depends only on the 
value of $S_i$ at that atom, this
allows us to minimise $U_{tot}$ with respect to $\{S_i\}$ 
at each site {\it independently} which gives us:

\begin{equation}
S_i=\frac{-2N}{W_i}\left [ I+\sum V_m(r_{ij})\right ]
\end{equation}

    Values of $S_i$ are bounded by 0 and 2N-T.  Thus, we can solve analytically for $S$, and the
computational cost of the calculation becomes equivalent to a standard
EAM.

The energy depends variationally on the values of $S_i$: $\partial
U/\partial S_i=0$.  This allows application of the Hellmann-Feynman
theorum in the force calculation to eliminate terms involving changes
in $S_i$.

Thus the complexity of the forces is also just like EAM

\begin{eqnarray}\label{eqn:HF}
F_i &=& \sum_j \frac{1}{2} \frac{dV_0}{dr_{ij}} + S_i\frac{dV_m}{dr_{ij}} {\bf \hat{r_{ij}}}
\\ 
&+&\left(\frac{S_i^2+T^2}{4N}-\frac{T}{2}\right) \sum_j\frac{d\phi}{dr_{ij}}\hat{\bf r_{ij}}
\\
&+&\sum_j\left(\frac{S_j^2+T^2}{4N}-\frac{T}{2}\right) \frac{d\phi}{dr_{ij}}\hat{\bf r_{ij}}
\end{eqnarray}

To make a potential, one needs to fit the pairwise functions
$V(r_{ij})$ and $\phi(r_{ij})$ and the parameter $I_0$.  This model can be
generalised from the Finnis-Sinclair form to the EAM form by using a
parameterised embedding function rather than the square root.

\begin{figure}[htb]
\centerline{\scalebox{0.6}{
\includegraphics{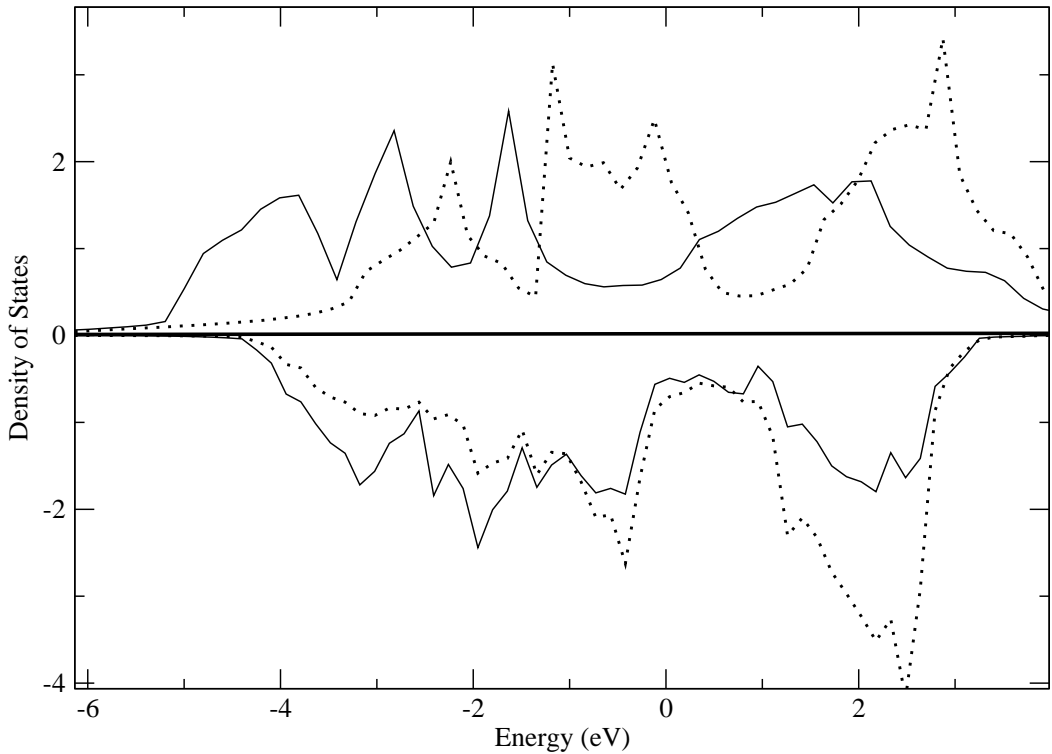}}}
\caption{\label{fig:MoV} Density of states from ab initio GGA
pseudopotential calculations with 4913 k=points, adjusted so that the
Fermi energy lies at the zero of energy.  Upper bands: pure V (dots)
and Mo (solid) at their equilibrium structure (bcc, 3.006\AA and
3.184\AA respectively).  Lower bands (plotted upside down for ease of comparison): MoV alloy in CsCl structure
(lattice parameter 3.086\AA), with density of states projected onto
the two atoms according to the integrated electron density within 
1.5\AA of the nucleus.  Further 16-atom supercell calculations for single 
substitutional  Mo in V and Cu in Fe show similarly undifferentiated bands, 
although when semi-core  $p$-electrons are included in the valance band this method show near-perfect localisation.
}
\end{figure}

\section{Two band model for alloys}
\subsection{Transition metal alloys}

An alternative use of a two d-band model occurs when an alloy of two
d-metals is formed.   The difficulty with one band on each atom
 is that charge transfer between sites 
becomes non-local and minimising energy becomes non-analytic\cite{skr} 
.  It is 
possible to use a fictitious dynamics for the charge transfer, in the same
spirit as Car-Parinello, but this would introduce a large number of additional equations of motion.

The convenient solution, implicit in equation  \ref{eq:MB}, is to assume
that a similar $d-$band can be projected onto both sites, shifted for charge neutrality.  This would allow us to use the MB approach.  Here we investigate
whether such an assumption can be justified.  Fig.\ref{fig:MoV} shows a
density of states for a MoV alloy.  The notable feature is that the
alloy bands are narrower than the pure metal bands, suggesting that
$2\phi_{AB}(r_{ij}) < \phi_{AA}(r_{ij})+ \phi_{BB}(r_{ij})$

Furthermore, calculations of the projected density of individual
electron states onto individual atoms for isolated Mo and V and for 
Cu in Fe shows that there are no localised $d$-states - the band 
structure is projected fairly evenly onto each species.

\begin{figure}[htb]
\centerline{\scalebox{0.6}{
\includegraphics{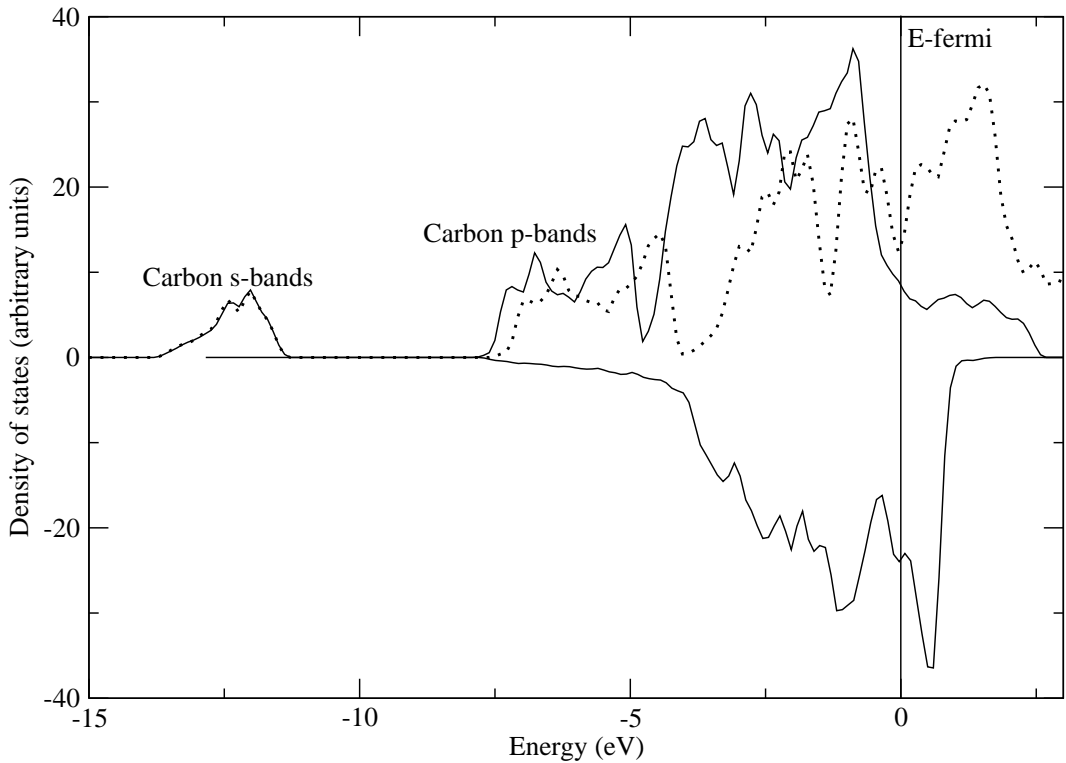}}}
        \caption{\label{fig:Fe3C} Density of states from ab initio GGA
pseudopotential calculation with 1000 k=points.  Upper bands: Fe$_3$C
cementite with atomic positions and unit cell are fully relaxed
(a=5.069\AA, b=6.744\AA, c= 4.514\AA).  Lower bands: unit cell
containing Fe in the same locations as cementite, but no carbon, this
gives a non-ferromagnetic ground state.
}
\end{figure}
 
\begin{figure}[htb]
\centerline{\scalebox{0.6}{
\includegraphics{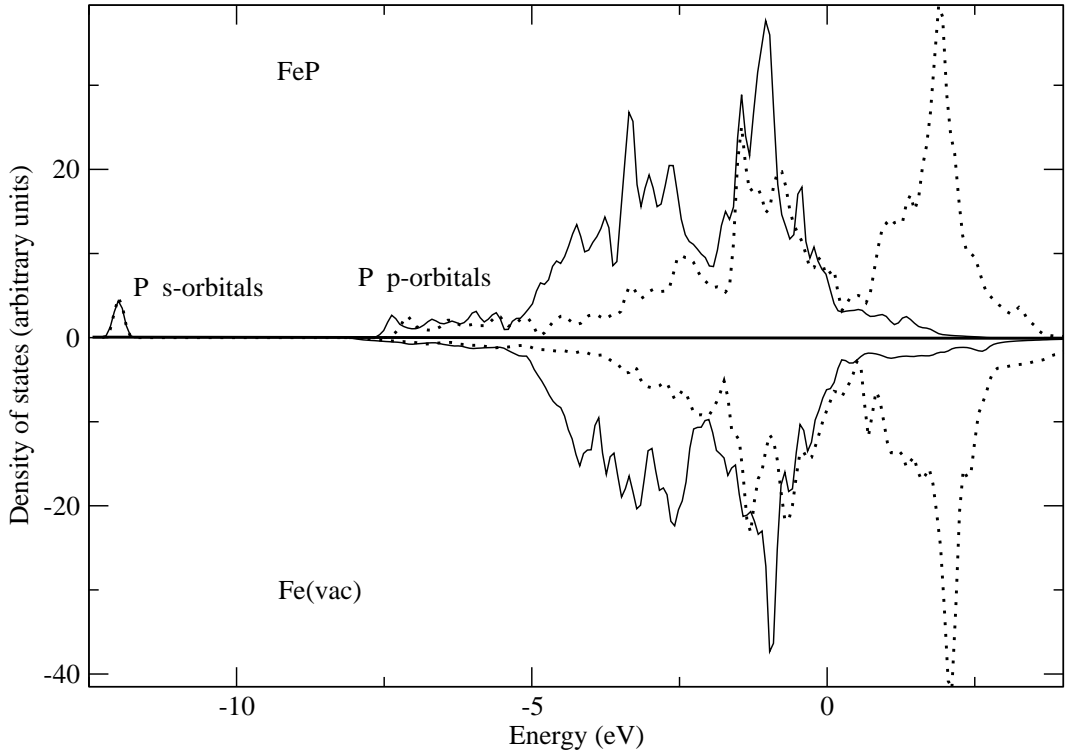}}}
\caption{\label{fig:FeP} Density of states from ab initio GGA
pseudopotential calculations with 729 k=points, adjusted so that the
Fermi energy lies at the zero of energy.  Upper bands: supercell of 15
iron atoms with a substitutional P impurity, relaxed to its
equilibrium volume (186$\AA^3$) with ions relaxed and magnetic moment
35.3$\mu_B$.  Lower bands:supercell of 15 iron atoms with a vacancy,
relaxed to its equilibrium volume (183.7$\AA^3$) with ions relaxed and
magnetic moment 36.4$\mu_B$.  Note that the two densities of state are
extremely similar, the main effect of the P being the addition of two
low lying s-states and some additional $p$-states and the low energy
end of the $d$-band.
}
\end{figure}

\subsection{p-band impurities - C, P, Al}

The presence of non-transition metal impurities in steels is of major
interest for a wide range of applications.  The combination of p- and
d-based materials makes for a complicated band structure and the picture
of simply adding electrons to a rigid band is no longer tenable.
Fig.\ref{fig:Fe3C} shows the ab initio calculated density of states for
Fe$_3$C cementite, compared with the density of states calculated for
the identical arrangement of Fe atoms with the carbons removed.

The first striking feature is that the occupied carbon bands can be
readily distinguished in Fe$_3$C. Unoccupied carbon $p$-bands lie well
above the Fermi energy, meeting the requirement for charge neutrality
which underlies the idea of and band energy determined from a 
local density of states\cite{avf}.  This suggests
that, in the FS picture at least, the hopping integral between bands
can be ignored.  Furthermore, the carbon bands lie far below the Fermi
energy.  This suggests that they will not contribute to the metallic
bonding and cannot be treated in the same way as the $d$-electrons.
This means that to a first approximation one can set $\phi_{FeC}(r) =
0$, and treat the iron-carbon interaction purely as a pair potential.

Similar unhybridised carbon states have been seen for isolated
defects: the one occasion when the bonding of the
dissolved carbon changes qualitatively being the formation of a covalent bond
in a C dimer located in a Fe vacancy\cite{domain2}.

A similar same situation applies to embedding a phosphorus atom in iron.
The phosphorus $s$-electrons play no part in the bonding, and the 
$p$-states lie below the $d$-band (Fig.\ref{fig:FeP}).

\section{Multiple elements without refitting}

Finnis-Sinclair type models have had reasonable success in modelling
interatomic alloys of transition metals, here the extension of the
theory is straightforward.  If accurate results for radiation damage
are required, then some refitting is necessary.  For alloys with low
concentrations of impurities, such as reactor steels, a more
straightforward approach may be appropriate.  With the simple 
two-band iron model above, we can assume that the main effect 
of small concentrations of local impurities is not to change 
the shape of the projected band structure, but to change the number of 
electrons in the band.  As before, the MB energy can be separated 
into contributions from each atom:

\begin{equation} U = \sum_i U_i^\uparrow + U_i^\downarrow + U_i^{XC} = +
\frac{W}{4N}(T_i^2+S_i^2) - T_iW/2 - E_{i0}S_i  \end{equation}

following the procedure adopted for Cs\cite{skr} an approximate
potential for an embedded transition metal impurity  
can be made using the same parameters as for iron with lengths scaled by the Fermi vector
($T^{-1/3}$) and energies by $T^{1/2}$.  The separation of the 
pair potential into the form:

\begin{equation} V_{IJ}(r_{ij}) = V_I(r_{ij}) + V_J(r_{ij})
\end{equation}

implies that the screening for the impurity is due to  {\it
the iron charge density}. Hence such a pairwise potential would not be
transferrable to environments of high concentration of the impurity
atom.  It is already well known that potentials fitted to give a good
description of the dilute environment show poor transferrability
across the phase diagram.\cite{Caro,Calder}

\section{Potentials for molecular dynamics - or vice versa}

The concept of designing interatomic potentials is typically to 
start with a full quantum mechanical treatment of the electrons, 
and then via a series of approximations arive at a form suitable 
for use in molecular dynamics.  Hence EAM is formulated as a 
extreme localised form of the density functional theory, while 
Finnis Sinclair is a simplified tight binding.  It is unclear to 
what extent the correctness of the original physics is retained 
- certainly in  applications empirical refitting is used 
to "fold back" all the missing physics in an approximate way.

The ultimate test of the usefulness of a potential is its 
transferrability - does it work in environments different 
from where it was fitted?  If it is transferrable, it is quite acceptible for 
a potential to include arbitrarily introduced terms to compensate for the
uncontrolled approximations made from the full quantum mechanical treatment.

Hence an alternate view of potentials is to {\it start} with the needs
of molecular dynamics and ask what functional forms are possible.  The
constraints are that we require short-ranged interactions based on
atomic positions only.  The potential may then use any
information available from the molecular dynamics to reproduce the
forces correctly.  In fact, much information is available in a real
molecular dynamics calculation which might be useful in formulating
the potential.

For simplicity, we assume the calculation does not explicitly evaluate
three-body terms - this involves an additional calculational overhead
and there are no such terms in the full many-body quantum mechanical
Hamiltonian.  The
following quantities can be evaluated:

a) functions of separations at each atom (EAM's 
"electron density" or FS's "sum of squared hopping integrals").

b) functions of other rotationally invariant quantities at each atom (as in MEAM)\cite{Baskes}.

c) on-site parameters with respect to which the energy can be minimised {\it locally} and 
{\it analytically} (as in the two-band model\cite{skr}).

d) the mean electron density, and the free electron Fermi energy.

\subsection{Evaluating and Using the Fermi energy}

Almost all quantum treatments of metals require calculation of the
Fermi energy. In the Finnis Sinclair approach it is the fixing of the 
Fermi level combined with the condition of charge neutrality which allows the method to work for other than half-filled bands.\cite{avf} 
Further, it has been known since Hume-Rothery that
Fermi surface effects lies at the heart of the structural properties
of metallic alloys. 

In principle, a short-ranged potential cannot
incorporate such effects, because the Fermi energy depends on
delocalised electrons: an extremely long range effect.  However, 
in setting up a
molecular dynamics simulation, all the information needed to calculate
the Fermi level is often included.  Most radiation damage calculations
entail a constant volume ensemble simulation of a bulk material with a
few defects: the mean electron density is known and constant
throughout, and the Fermi level can therefore be established.  In
surface calculations there appears to be a problem, however even here
the molecular dynamics uses boundary conditions (periodic slab
geometry, fixed layer, etc. ) which are intended to represent contact
with a bulk of fixed density.  Again, the Fermi Energy for this can
readily be established.  In constant pressure simulations, it may
appear that the Fermi energy should vary, yet even this is unclear -
in standard Parrinello-Rahman dynamics an expansion of the region of
interest is accompanied by and expansion of all periodic-image
supercells.  In the case of a phase transition, this is reasonable and
the Fermi energy can track the density, however in other cases the
local expansion would actually cause a compression in the surrounding
region, preserving the Fermi energy at its bulk value.  Hence either
treatment of the Fermi level has equally good physical justification
as the Parrinello-Rahman scheme itself.

Thus there should be no objection to using the Fermi Energy as part of 
the formulation of a potential, since its value can readily be determined 
at the outset of a molecular dynamics calculation.

One anomaly this removes is the series of phase transitions under
pressure exhibited by empirical potentials: e.g. even the 0K ground state
of the Lennard-Jones potential switches between
fcc and hcp as a function of pressure\cite{anj}, though no real material does
this.  Other close-packed
metal potentials behave similarly.  The reason for this is that phase
stability depends on the long-range part of the potential - beyond
second neighbours at equilibrium.  Under compression various shells of
neighbours come into the range of the potential, increasing or
decreasing the energy of the two phases.

The resolution to this problem has long been known.  Thirty-five years ago 
Heine and Weaire\cite{hw} showed that within pseudopotential theory phase 
stability for simple metals the energy could be calculated as a volume term 
independent of atomic positions, plus a pair potential with asymptotic form:

\begin{equation} V(R) \propto \frac{\cos(2k_FR+\phi)}{(2k_FR)^3}  \end{equation}

The key point here is that $k_FR$ is close to constant for isotropic
compression.  Hence this term will not contribute to the change in
relative phase stability under pressure.  The $r^{-3}$ form can be
made usable in molecular dynamics if the long range pairwise part is
damped with an exponential\cite{PettiforWard}: this corresponds to a
finite electron temperature and does not significantly affect the
forces.  The same fictitious-electron-temperature approximation is
used routinely in {\it ab initio} calculations where it is known as
"Fermi smearing".

Recently, potentials which attempt to measure the density $\rho_i$
with local sampling have appeared\cite{FWG}: the most
effective local sampling measure seems to be a sum of a Gaussian
potential. These allow the pair potential to depend explicitly on the
electron density and take the form:

\begin{equation}U = \sum_{ij} V_{I}(r_{ij},\rho_i) 
+ \sum_i F_I[\sum_j \phi_{IJ}(R_{ij})] \end{equation}

However, as discussed above, for practical molecular dynamics such short range
approximations to find the Fermi energy (and their associated calculational cost) are
unnecessary.

\section{Conclusions}

In this paper we have laid out how the ideas underlying the two-band
model, already applied to caesium\cite{skr}, can be extended to
magnetic elements such as iron and its alloys.  By way of {\it ab
initio} simulation we have shown how the site-projected density of
states, which forms the conceptual framework behind Finnis-Sinclair
type potentials, suggest particular approaches to different systems.
In particular, if magnetism is introduced without affecting the
pairwise potential, an EAM-type model is recovered - the onsite
magnetisation need only be calculated if different Pauli repulsion for
majority and minority bands is assumed.  Similarly, for sp-bonded
impurities C and P in iron, we find valence states well below the
Fermi energy and little effect on the d-bands.  This suggests that
these impurities can be sensibly treated with pair potentials + unmodified 
many-body iron potential, unless they get close enough to bond to one 
another.

Accurate parameterisations are not presented here, rather the paper
points the way to developing future interatomic potentials and
provides some {\it post facto} justification for previous studies.


\begin{thebibliography}{20}
\bibitem{rose} JH Rose JR Smith F Guinea J Ferrante Phys. Rev. B 29, 2963 (1984)
\bibitem{thet} GJ.Ackland and R.Thetford, Phil.Mag.A, 56, 15. (1987).
\bibitem{SuengWu} SW Han
\bibitem{db} MS Daw, and MI Baskes,  Phys.Rev.B  29, 6443 (1984)
\bibitem{HKS} P. Hohenberg W. Kohn and L. J. Sham  {\it   Phys. Rev.}   \textbf{136} B864 (1964)
\bibitem{FS} MW Finnis and JE Sinclair, Phil.Mag.A,  50, 45. (1984)
\bibitem{duc} F. Ducastelle, {\em J. Phys. Paris} {\bf 31}, 1055 (1970)
\bibitem{PettiforBook} DG Pettifor in ``Electron Theory in alloy design''
eds DG Pettifor and AH Cottrell IoM, London p81 (1992)
\bibitem{bop} DG Pettifor Phys. Rev. Lett. 63, 2480 (1989)
\bibitem{Baskes}MI Baskes, {\it Phys.Rev.B} {\bf 46} 2727 (1992)
\bibitem{mendel} MI Mendelev, SW Han, DJ Srolovitz, GJ Ackland, DY Sun. and M Asta, Phil. Mag. A, {\bf 83}. 3977 (2003)
\bibitem{domain1}    C. Domain and C. S. Becquart Phys. Rev. B 65, 024103 (2002)
\bibitem{domain2}    C. Domain, C. S. Becquart and J.Foct Phys. Rev. B 69, 144112 (2004) 
\bibitem{skr} S.K.Reed, G.J.Ackland Phys.Rev.B {\bf 67}, 174108 (2003)
\bibitem{avf} G.J.Ackland, V.Vitek and M. W. Finnis, {\em J. Phys. F} {\bf 18}, L153. (1988)
\bibitem{Caro} EM Lopasso, M. Caro, A.Caro and PEA Turchi {\it Phys.Rev.B} {\bf 68} 214205 (2003)
\bibitem{Calder}G.J.Ackland,
D.J.Bacon, A.F.Calder and T.Harry Phil.Mag.A, 75 713-732 (1997)
\bibitem{hw} V. Heine and D.Weaire Solid State Physics {\bf 24} 249 (1970)
\bibitem{PettiforWard} DG Pettifor and MA Ward, Solid State Commun., 49 (1984) 291.
\bibitem{FWG} MW Finnis, AB Walker and P Gumbsch J.Phys.: Condensed Matter, 10, 7983-7993, (1998) 
\bibitem{anj} Jackson AN, Bruce AD, Ackland GJ  Phys Rev E 65 (3) 036710 (2002)
\end{thebibliography}
\end{document}